\documentclass{article}
\usepackage{latexsym}
\usepackage{amsmath,amssymb}
\usepackage{float}
\usepackage[dvipdfmx]{graphicx}%

\begin{document}

\pagestyle{plain} 
\setcounter{page}{1}
\setlength{\textheight}{650pt}
\setlength{\topmargin}{-40pt}
\setlength{\headheight}{0pt}
\setlength{\marginparwidth}{-10pt}
\setlength{\textwidth}{20cm}

\title{A study of Inverse Ultra-discretization of  cellular automata}
\author{Norihito Toyota  \and Faculty of Business Administration and Information Science, \\ \hspace{6mm} Hokkaido Information University, Ebetsu, Nisinopporo 59-2, Japan\\
toyota@do-johodai.ac.jp }
\date{}
\maketitle
%
%



\abstract{
 In this article, I propose a systematic method for the inverse ultra-discretization of cell automata using a functionally complete operation. 
We derive  difference equations for the 256 kinds of  elementary cellular automata(ECA) introduced Wolfram\cite{wolfram}  by the proposed means of the inverse ultra-discretization.   
We show that the behaviors of  ECAs can be completely reproduced by numerically solving the obtained difference equations. 
\subsubsection*{keyword; elementary cellular automaton, functionally complete operation, \\ \hspace{17mm} inverse ultra-discretization, difference equation}


\section{Introduction}
It is known that we can obtain difference equations by discretizing independenent variables, time and space, of differential equations. 
Moreover the method of the ultra-discretization has been established \cite{gibo}\cite{Iwa}\cite{Tomo} where the dependent variable is also discretized 
to get cellular automata  (CA)\cite{Il}. 
This method has mainly developed in a field of traffic flow. 
For an example, We can derive the descritized Burgers' equation by descritizing the Burgers' equation. 
Moreover we obtain the ultra-discretized  Burgers' equation by descritizing the dependent valiable of it\cite{Nishi}, which corresponds to rule 184 ECA\cite{Worfram}.
Recently the method of the inverse ultra-discretization has developed\cite{eca}. 
Some CAs are lead to the corresponding difference equations by an inverse process of the ultra-discretization. 

In this article,  we propose a systematic method to inversely ultra-discretize CA by using a functionally complete operation, for example, 
And-gate, OR-gate and NOT-gate.  
To inversely ultra-discretize CA, the method using local linear equations and Max Plus algebra are developed in as ever\cite{eca}. 
In such a method, it is not clear how to derive the local linear equations. 
It is not seemed that there is any general algorithm in the sutudies. 
On the other hand, Beniura and Nakano have proposed a method to  inversely ultra-discretize ECAs systematically by studying CAs connected in some transformations\cite{kure}. 
This method, however, is not clear to  work in other CAs except for ECA. 
The method proposed in this article is systematic and solid because the method is based on functionally complete operations. 
Using this method, I inversely ultra-discretize 256 kinds of ECA introduced by Worfram\cite{Worfram}. 
 Then I numerically solve the difference equations given by the inverse ultra-discretization and show that the behaviors of the solutions reproduce the properties of the original  ECAs wholly .
 
 \section{Inverse ultra-discretization of CA}
 In this article, $t$ represents a discrete time, $n$ is a position of a cell and  $u \in \{0,1\}$ is a state variable. 
 Thus $u_n^t$ represents a sate in $n$-th cell at time $t$.
 A presented method of the inverse ultra-discretization is described below. \\
 
 1. derive a logical expression consist of ANDi$\wedge$)COR($\vee$ ) and NOT($\bar{u}$ ) which reproduces the outputs of a CA
 
  For example, focusing the following  rule of a CA with 2 states and 3 neighborhoods
  $$u_{n-1}^t \; u_n^t \; u_{n+1}^t = 101 \rightarrow u_{n}^{t+1}=1,$$
  we obtain the following logical expression;
$$ u_{n-1}^t \wedge \overline{u_n^t} \wedge u_{n+1}^t.$$
Thus we derive logical expressions for  all outputs of the CA and connect their logical expressions by OR (principal disjunctive canonical form).\\

2. rewrite the constructed logical expressions using AND and NOT as a sort of a minimal functionally complete operation. 
 
By using the de Morgan's theorem, OR is rewritten by
\begin{equation}
   x \vee y=  \overline{\bar{x}\wedge \bar{y}}.
\end{equation}
The following formula is also useful for the exclusiveOR on occasion. 
\begin{equation}
   x \oplus y=   \overline{x \wedge y}   \wedge \overline{\bar{x}\wedge \bar{y}}.
\end{equation}  

3. transform the presented logical formula to Max operation.\\
 First the following transformation  is made for the presented formula consist of AND and NOT;
\begin{eqnarray}
\overline{u_n^t} &\longrightarrow& (1-{u_n^t}), \nonumber \\
 u \wedge v &\longrightarrow& u\times v,
\end{eqnarray}
where $\times$ represents a usual multiplication. 
Then we set up a formula max[$F(u_{n+1}^t,u_n^t,u_{n-1}^t,0$] for the constructed formula $F(u_{n+1}^t,u_n^t,u_{n-1}^t)$.\\

4. construct a difference equation by using the following  ultra-discretization formula for real numberes $A$ and $B$; 
\begin{eqnarray}
			\max
			\left[
				A,B
			\right]
			=
			\lim_{\varepsilon \to 0} \varepsilon \log
			\left[
				\exp
				\left(
					\frac{A}{\varepsilon}
				\right)
				+
				\exp
				\left(
					\frac{B}{\varepsilon}
				\right)
			\right], 
		\end{eqnarray} 
where the resltant formula is written as $u_n^{t+1}=E_{\varepsilon (no. of rule)} (u_{n-1}^t, u_n^t, u_{n+1}^t)$. 	

Really we fix $\varepsilon$ to a small positive finite value, because we cannot actually take $\varepsilon\rightarrow 0$ on a computer. 
Then $u_n^t$, however, converges to $0$ after many time steps and so the resultant difference equation cannot support the properties of ECA. 
 We need the following step 5 in order to avoid this distress. \\
 
 5. set up simultaneous difference equations by using a kind of filter\cite{eca}. 
 
 First of all, replace the derived formula $u_n^{t+1}=E_{\varepsilon (no. of rule)} (u_{n-1}^t, u_n^t, u_{n+1}^t)$ for $u^{t+1}_n$ with $v^t$ definrd by 
 \begin{equation}
v_n^t = F_{\varepsilon (no. of rule)} (u_n^t)=\frac{1}{1+e^{-(u_n^t -\Delta)/\varepsilon} },
 \end{equation}
 where $\Delta$ takes a very small positive value as $\varepsilon$.  
 From (5), $F_{\varepsilon (no. of rule)} (u_n^t)$ comes close to $0$  when  $u_n^t -\Delta<0$
  and $F_{\varepsilon (no. of rule)} (v_n^t)$ comes close to $1$  when $u_n^t -\Delta>0$. 
  Thus $\Delta$ paly a role of a threshold. 
 
 \section{Example}
 We explain the presented method in the case of rule 161 among 256 ECAs as an example in this section. 
 The transition rule of the rule 161 ECA is given by Table 1. 

 	\begin{table}[h] \caption{Rule 161}
 	 \begin{center}
		\begin{tabular}{|c|c|c|c|c|c|c|c|c|} \hline
			$u_{n-1}^t,u_n^t, u_{n+1}^t$ & 1 1 1 & 1 1 0 & 1 0 1 & 1 0 0 & 0 1 1 & 0 1 0 & 0 0 1 & 0 0 0  \\ \hline
			$u_n^{t+1}$ & 1 & 0 & 1 & 0 & 0 & 0 & 0 & 1\\ \hline 
		\end{tabular}
     \end{center}
	\end{table} 

For simplicity, We use the logical formula (6) given by Wolfram\cite{wolfram} instead of the way given in the example in Step 1 of the previous section(Step 1). 
We can show that the way given in the example also gives essentially the same results as the method proposed here.	
	
	\begin{equation}
		u_n^{t+1} = u_{n-1}^t \oplus u_{n+1}^t \oplus (u_{n-1}^t \lor u_{n+1}^t \lor \overline{u_n^t}).
	\end{equation}	

Using (2), 	
we rewrite this logical formula with one represented by AND and NOT(Step 2).
\begin{eqnarray}
	u_n^{t+1} = 
		\left(
			\overline{	
				\overline{
					u_{n-1}^t
				}
				\land
				\overline{
					u_{n+1}^t
				}
				\land
				\overline{
					u_{n-1}^t
				}
				\land
				\overline{
					u_{n+1}^t
				}
				\land
				u_n^t
			}
		\right)
		\land 
		\left(
			\overline{
				\overline{
					u_{n-1}^t
				}
				\land
				u_{n+1}^t
				\land
				\left(
					\overline{
						\overline{
							u_{n-1}^t
						}
						\land
						\overline{
							u_{n+1}^t
						}
						\land
						u_n^t
					}
				\right)
			}
		\right)
		\land \nonumber \\
		\left(
			\overline{
				u_{n-1}^t
				\land
				\overline{
					u_{n+1}^t
				}
				\land
				\left(
					\overline{
						\overline{
							u_{n-1}^t
						}
						\land
						\overline{
							u_{n+1}^t
						}
						\land
						u_n^t
					}
				\right)
			}
		\right)
		\land 
		\left(
			u_{n-1}^t \land u_{n+1}^t \land
			\overline{
				u_{n-1}^t
			}
			\land
			\overline{
				u_{n+1}^t
			}
			\land
			u_n^t
		\right).
	\end{eqnarray}

We apply the transformation (3) to (7) and set up a formula using max-operation(Step 3). 
So we obtain the following defference equation from (7).
		\begin{eqnarray}
		u_n^{t+1} = \max[
			\Bigl(
				1-
					(1-u_{n-1}^t )
					\times
					(1-u_{n+1}^t )
                   \times   
                    (1-u_{n-1}^t  )\times  
					\left(
							1-u_{n+1}^t
						\right)
						\times u_n^t
				\Bigr)
						\times   \Bigl(1-	 
&&		 €\nonumber \\ 
              \left( 1-u_{n-1}^t \right)	
     	  \times u_{n+1}^t        \times
                 \bigl(1-                      
                         \left(
							1-u_{n-1}^t
						\right)
						\times								
 \left( 1-u_{n+1}^t  \right)		 \times 	u_n^t
			    \bigr) \Bigr)  \times    \Bigl(1-
				u^t_{n-1} \times \left(1-u_{n-1}^t \right) \times
&&  €\nonumber \\ 	 
\bigl(1- \left(1-u_{n-1}^t	\right)  \times \left( 1-u_{n+1}^t \right)	\times	u_{n+1}^t
	\bigr) \Bigr)				\times
		\Bigl(1-						
u_{n-1}^t \times	 u _{n+1}^t  \times  \left( 1-u_{n-1}^t  \right)	\times  	
      \left(	1-u_{n+1}^t   \right)
                    \times 
							u_n^t
		\Bigr)	, 0
		],
	\end{eqnarray}
	where we interpret $u_n^t$ as a real number	$u_n^t\in[0,1]$. 

Moreover we obtain the following defference equation by using the ultra-discretization formula (4) (Step4);		
	\begin{eqnarray}
		u_n^{t+1}&=&E_{\varepsilon (161)}
			\left(
				u_{n-1}^t, u_n^t, u_{n+1}^t
			\right),  \\ 
		E_{\varepsilon(161)} 	\left(
				x, y, z \right)
			&\equiv & \varepsilon \log
			\left[
				\frac{1}{2}
					\exp\left\{
						\Bigl( 1-(1-x)  \right.  \right. 
  \times (1-z)(1-x)(1-z)y \Bigr)   \exp \Bigl( 1- (1-x) 
\nonumber \\ 
&&			\times z(1-(1-x)(1-z)y) \Bigr)   \exp \Bigl( 1-x)     
				\times (1-z)  ((1-(1-x)(1-z)y )\Bigr) 
 \nonumber \\ 
 &&\exp \Bigl( 1- \left. 
			xz (1-x)(1-z)y ) \Bigr)/\varepsilon \right\} +1/2 ], 
		\end{eqnarray}
where we use the following notations; $x=u_{n-1}^t$, $y=u_n^t$, $z=u_{n+1}^t$. 
	
$u_{n}^{t+1}$ comes close to $0$ with time steps as mentiond above. 
This behavior is shown in the left side of Fig.1. 

Thus we introduced the filter as explained in the step 5 and obtain (Step 5)  
		\begin{eqnarray}
			F_{\varepsilon (161)}(u) \equiv \frac{1}{1+\mathrm{e}^{-(u-\Delta)/\varepsilon}} && \\
			\left\{
				\begin{array}{lll}
					u_n^{t+1}&=&E_{\varepsilon (161)} 
					\left(
						v_{n-1}^t, v_n^t, v_{n+1}^t
					\right)  \\
					v_n^t&=&F_{\varepsilon (161)}
					\left(
						u_n^t
					\right).
				\end{array}
			\right.
		\end{eqnarray}
		
\begin{figure}[h]
\centering
	\includegraphics[width = 9.8cm]{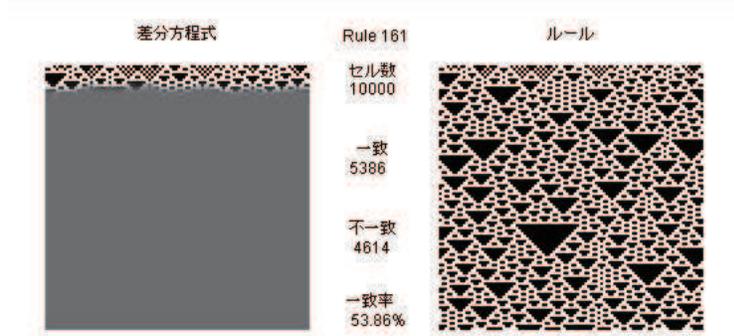}
	\caption{Behaviors of Rule 161 CA(right) and the solution of the corresponding difference equation without the filter(left).}
\end{figure}
\begin{figure}[h]
\centering
	\includegraphics[width = 9.8cm]{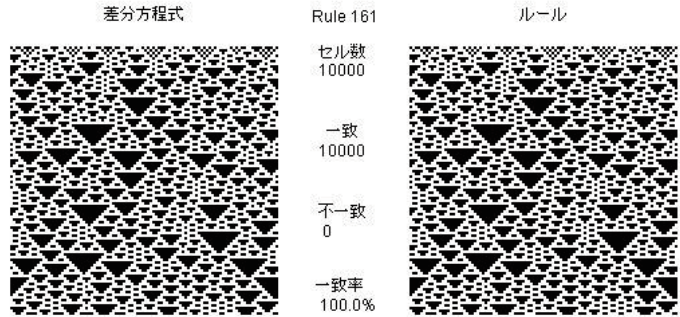}
	\caption{Behaviors of Rule 161 CA(right) and the solution of the corresponding difference equation without the filter(left).}
\end{figure}

\section{Numerical Results}
We numerically solve the difference equations obtained by the proposed procedure  under a periodic boundary condition. 
The behaviors of the numerical solutions are shown on a lattice, where the vertical direction represents time axis.  
The black cells show the state to be $1$ and the white ones show the state to be $0$. 
The initial state is taken to be homogeneously random, the size of the cell number (horizontal size) is $1000$ and it takes $1000$ time steps.   

The right side of Fig.1 shows time developments for rule 161. 
The left side of Fig.1 shows  the numerical solution of (10) without filter where the behabvior is almost same as one of the right figure up to a few steps,  
but the states of all cells rapidly converse to $0$-state after some time steps. 

The left side of Fig.2 shows the numerical solution of (11) and (12) with the filter. 
Compared with the right of Fig.2 which shows the result of rule 161, we observe that both results agree perfectly. 

We carried out these procedures for all ECAs and confirmed that  the behaviors of ECAs and those of the solutions of the corresponding difference equations 
are completely same for in all cases.
These observations show that our method of the inverse ultra-discretization is correct. 

\section{Conclusions}
I propose a systematic method for the inverse ultra-discretization via the logical formulas, especially a functionally complete operation, in this article. 
It is characteristic to adopt a minimal functionally complete operation, AND and NOT in the presented method.
So it guarantees that this procedure stands good for all CAs with 2 states.  
We carried out our procedure for all ECAs and confirmed that the behaviors of ECAs and the solutions of difference equations completely agree in all cases.
This fact justifies our method of the inverse ultra-discretization. 

For inversely ultra-discretizing CA, the method using local linear equations and Max Plus algebra have been studied in a conventional manner\cite{eca}. 
In such method, it is not clear how to derive the local linear equations. 
Any general algorithm to get local linear equations seems not to be given in the sutudies. 
Beniura and Nakano have proposed a systematic method to  inversely ultra-discretize CA  by studying CAs connected in some transformations\cite{kure}. 
This method is not clear to  work in other CAs except for ECA. 
Their method leads to defference equations that differ from the ones in this article, but the behaviors of numerical solutions of the both difference equations 
are the same for time developments.  
This fact shows that there can be several corresponding difference equations for a CA. 
This may be effective to study the equivalence class of difference equations. 
Furthermore to derive corresponding partial differential equations from the defference equationsis is worth studing.

\end{document}